# Infrared Computer Vision for Utility-Scale Photovoltaic Array Inspection

David F. Ramirez[1], Deep Pujara[1], Cihan Tepedelenlioglu[1], Devarajan Srinivasan[2], Andreas Spanias[1]

[1]SenSIP Center, School of ECEE, Arizona State University, Tempe, AZ 85281, USA

[2]Poundra, LLC, Tempe, AZ 85281, USA

[1]{dframire, dpujara1, cihan, spanias}@asu.edu, [2]srini@poundra.com

*Abstract*—**Utility-scale solar arrays require specialized inspection methods for detecting faulty panels. Photovoltaic (PV) panel faults caused by weather, ground leakage, circuit issues, temperature, environment, age, and other damage can take many forms but often symptomatically exhibit temperature differences. Included is a mini survey to review these common faults and PV array fault detection approaches. Among these, infrared thermography cameras are a powerful tool for improving solar panel inspection in the field. These can be combined with other technologies, including image processing and machine learning. This position paper examines several computer vision algorithms that automate thermal anomaly detection in infrared imagery. We demonstrate our infrared thermography data collection approach, the PV thermal imagery benchmark dataset, and the measured performance of image processing transformations, including the Hough Transform for PV segmentation. The results of this implementation are presented with a discussion of future work.**

*Keywords*—**photovoltaic system, solar energy, solar panels, infrared imaging, image processing, computer vision, machine learning, object detection, infrared thermography**

## I. Introduction

Utility-scale solar panel arrays provide a desirable renewable energy solution; however, large-scale photovoltaic (PV) energy has unique operational challenges. For utility-scale, a PV array includes hundreds of solar panels in an array and, in some cases, thousands of panels across a generation plant, potentially producing up to a gigawatt of power at peak production [1-2]. Ensuring their long-term operation, uptime, generation efficiency, and cost-effectiveness can be difficult [4-7].

Individual PV panels are commonly affected by faults that degrade power generation performance. Firstly, identifying manufacturing defects rapidly is crucial, as these defects can affect immediate power generation and return on investment [8-9]. Although industrial-grade panels are built for ruggedness and longevity, damage due to shipping and installation can occur. Equally important is to identify damages that occur over time due to extreme weather events, ambient temperature, circuit issues, and other unforeseen external circumstances [10]. Detecting and correcting environmental factors such as transient partial shading [11-12] and soiling [12-13] from dust or bird droppings are essential for the overall efficiency of the PV array. Finally, PV cells are semiconductors that inherently degrade over time with heat and use due to doped silicon atomic migration [14-15]. It is beneficial to estimate and track the intrinsic silicon degradation, micro-cracks [9], and delamination [16] over the life of each panel. Severe damage can be a safety concern, causing short-circuit or open-circuit conditions. When a solar cell's electrical connection is open with no current flowing, the cell will be at a cool temperature.

All other types of damage and inefficiencies may cause a photovoltaic module to change from a power source into a power sink. The electrical energy of the array will be converted into heat energy spread across a solar panel, a string of panels, or as a hotspot in an individual photovoltaic cell [11]. This hotspot further deteriorates the affected region's overall power output and lifespan [17-18]. Any power generation inefficiencies will cause a disproportional degradation to the entire PV array circuit, with common parallel stings of panels degraded to the lowest common electrical DC voltage [19].

## II. Problem Statement

There is a need for sensors and algorithms to automatically detect the presence of damage, degradation, and faults to enhance the overall efficiency and safety of utility-scale PV arrays. Historically, solar inspection has relied heavily on visual and manual electrical inspections by a professional technician, a meticulous process of scanning panels for physical damage and electrical anomalies. While effective, this approach is costly, labor-intensive, and not well-suited for the scale of utility power plants. As a complement to regularly scheduled inspections, sensors and algorithms can automatically monitor solar-generated electrical measurements to detect and localize faults [19-23]. This real-time monitoring can automatically trigger alerts or maintenance, improving power generation and reducing downtime.

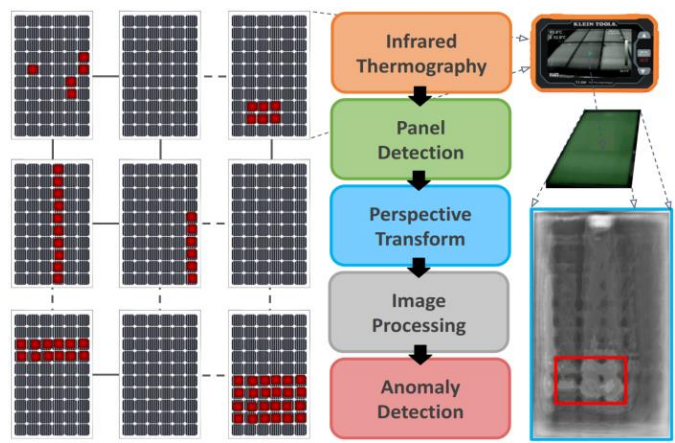

Fig. 1. Diagram depicts a PV array with hotspot anomalies, thermal imagery capture, data processing steps, and the data products produced by these steps.

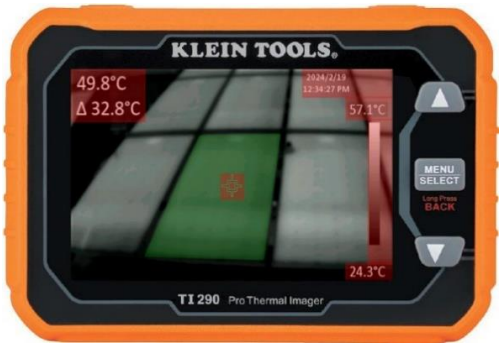

Fig. 2. An infrared thermography camera used for data collection [36]. The image shows masked overlay information regions in red and the most prominent PV panel in green.

Recent advancements have introduced innovative solutions for detecting faults, such as flying drones with high-resolution cameras [24-25]. These can rapidly scan large areas of utility-scale solar plants, collecting useful images for analysis. Innovations in utility-scale solar array inspection include using thermal cameras, which, coupled with appropriate imaging algorithms, can provide impactful results.

## III. RELATED WORK

### A. Thermal Imaging

Thermal imaging collected through infrared (IR) cameras has emerged [25-32] as a powerful technique for PV fault detection. These IR thermography cameras have recently become accessible to consumers and professionals due to advances in manufacturing and the relaxation of foreign military-use concerns. By detecting variations in the thermal image of a solar panel, these handheld tools can be used to identify hotspots caused by damage and degradation, allowing for targeted maintenance efforts. Professional technicians and electrical professionals use thermal cameras across many industrial applications, including the inspection of solar panels.

### B. Computer Vision

Researchers have proposed many unique algorithms for the automated detection of faults in the thermal imagery of solar panels. Some methods include image processing techniques to exaggerate hotspots in the greyscale infrared images. This may include pixel intensity regularization, contrast stretching or equalization, dilation to increase the size of hotspots, the Hough Transform to detect lines and shapes, edge detection, segmentation, masking and cropping of non-photovoltaic material, and finally, a detector of hotspot anomalies. Victor et al. [26] proposed a Hough transform-based detector to identify the location of a hotspot in an image. Using similar image processing methods, Moath et al. [25] developed a hottest spot segmentation method using an intensity difference threshold.

### C. Machine Learning

In addition to image processing techniques, researchers have implemented different machine learning (ML) techniques, such as statistical detectors, classifiers, and clustering methods for thermal images. To understand the complexity of these ML methods, it is helpful to understand the number of adaptive or trainable parameters in each method. Ngo et al. [27] utilized unsupervised ML clustering techniques, such as k-means and DBSCAN, to simplify the color range and group visually similar hotspot regions. This k-means method included 15 adjustable centroids on 1 numeric dimension, resulting in 15 trainable parameters. In other types of ML, an objective function is optimized using supervision with a known true value. Niazi et al. [28] implemented a Bayesian classifier to differentiate between defective and non-defective solar panels. Given five features as input and two possible outcomes, a Gaussian naive-Bayes method has 21 trainable parameters. Kurukuru et al. [29] again calculated intermediate image features but used a small artificial neural network (ANN) classifier instead. This ANN includes three fully connected layers of perceptrons and 176 trainable parameters to classify solar panels into eight categories.

### D. Deep Learning

Recent advances in ML, neural networks, and accelerated computing have enabled increasingly larger and more powerful methods, often called deep learning. Ksira et al. [30] utilized a convolutional neural network (CNN) to learn and classify spatial features. This method includes three convolution layers as a learned image filtering transformation and three fully connected layers for classification. This method uses 35,813 trainable parameters to predict PV panels as either five different faulty or operational categories. Instead of classification, Huang et al. [31] created a CNN-based solar panel edge-detection segmentation algorithm. This utilizes a customized CNN architecture, the MobileNet v2 backbone, as an image filtering transformation combined with a feature pyramid network (FPN) to generate an image-like segmentation mask. This method predicts a binary classification for each of the 320x256 output pixels and includes no less than 300,034 trainable parameters. Cong et al. [32] utilized a recent variety of the You Only Look Once (YOLO) object detection neural network architecture. This predicts the location of the bounding boxes around solar panels in an input image. This ML method is very large at 7,225,885 trainable parameters but is highly optimized for speed. With each new advancement, deep learning methods grow in size and complexity by an order of magnitude.

Automated analysis of thermal signatures, leveraging advanced image processing techniques and ML-trained predictors, can provide valuable insights into the health of PV arrays, enabling proactive maintenance and maximizing long-term energy production. Thermal cameras combined with automated detection algorithms enable rapid inspection of utility-scale solar plants, augmenting inspection procedures.

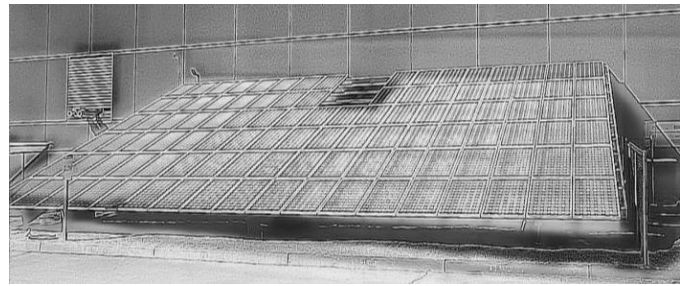

Fig. 3. Infrared thermography image of the solar testbed at ASU Research Park [1]. Image taken with the FLIR ONE Pro camera with multi-spectral dynamic imaging (MSX®) visual overlay added to low-resolution infrared.

## IV. METHODS

We present initial work on developing a new computer vision method for analyzing infrared image data of solar panels. Fig. 1 illustrates the system's data processing.

### A. Data Collection

The Arizona State University (ASU) Research Park includes a utility-scale solar array testbed and is the primary focus of the infrared photos [1]. Fig. 3 shows this solar testbed. This work utilizes the Klein Tools TI 290 infrared thermography camera. Fig. 2 shows the camera and an example image captured at a pixel resolution of 480x320 pixels. These images include overlay information from the camera used for calibration and measurements. We collect imagery of PV panels in standard operation, when not operating, and under faulted conditions such as soiling and shading [12, 23]. Some of the panels in our testbed also include signs of degradation, delamination, and cracks typical of weather damage and high temperatures. We collected 117 images of the utility-scale solar array testbed for use as a test set. These images include 755 PV array grid lines, 1050 array corner keypoints, and 900 PV panel bounding polygons, precisely labeled by hand.

### B. Image Processing

We perform several processing steps to detect, segment, isolate, normalize, and enhance the infrared temperature measurements of a PV panel. As a first step, we capture the pixel intensity-to-temperature scale for the measured high and low temperatures, as seen near the right bar in Fig. 2. These regions are then masked off, as they should not be included in the segmentation and measurements.

Next, the algorithm estimates the location of the most visually prominent solar panel. We use the Hough Transform to detect the edges of all visible PV panels. This maps out the grid pattern of the solar panels in the array. We evaluate the results of this edge and grid detection algorithm in Table 1.

With a quadrilateral polygon masking a PV panel in an image, we isolate only these image pixels and apply a perspective transformation to rectify the panel's dimensions.

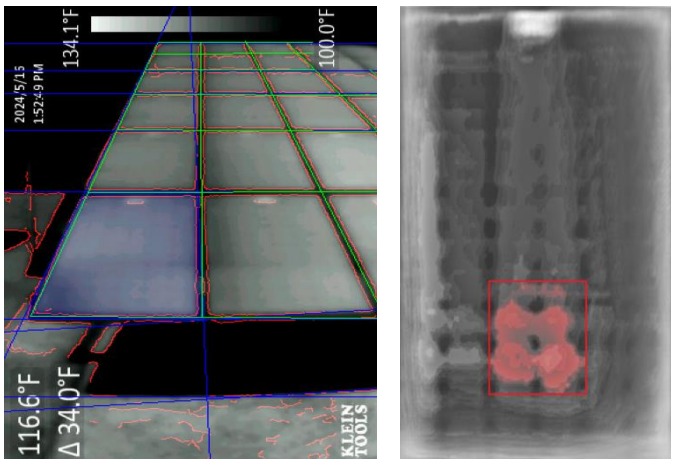

Fig. 4. Thermal image overlaid with detected red edge, blue grid prediction, and green ground truth lines. The proposed segmentation mask in blue is used to perspective-rectify the PV panel, and the pixel values are modified to show temperature variation. Suspected hot-spot PV cells are illustrated in red.

Next, the pixel intensity is also scaled to exaggerate the temperature differential. Fig. 4 shows an example of this data processing. We use a standardized contrast stretching method for consistency. Care must be taken to track the greyscale intensity changes relative to the temperature scale.

### C. Preliminary Results

We implemented a modified Hough Transform to detect the grid arrangement of PV panels in the utility-scale array. Our method applied a Gaussian blur, Canny edge detector, HoughLines [33] algorithm, and a custom method to group and average similar Hough line proposals. In some cases, multiple lines are proposed and matched to a single true line. In this case, only the multi-proposal counts as only a single true positive (TP). Table I measures the performance against the 755 grid line test labels.

TABLE I. PV ARRAY GRID LINE DETECTION

| Detection Result | Performance | | |
|---|---|---|---|
| | *Predicted* | *Equation* | *Metric* |
| True Positive | 463 | TP/755 | 61.3% Recall |
| Exceeds Error | 86 | Err/TP | 15.7% Error Rate |
| False Positive | 23 | TP/(TP+FP) | 95.3% Precision |
| False Negative | 206 | FN/755 | 27.3% Miss Rate |

## V. CONCLUSIONS AND FUTURE WORK

This position paper explored research published on infrared computer vision for PV array inspection. Utility-scale PV power plants are impacted by common solar panel faults, which can be observed as hotspots in thermal imagery. Algorithms that detect solar panels and hotspots, if present, can benefit the utility-scale inspection process.

Preliminary results demonstrate the opportunity and challenges of thermal imagery for PV. Detecting, segmenting, and isolating individual solar panels for analysis can be accomplished in many ways, each needing to be evaluated for benefits and shortcomings. Hotspot detection is a secondary task with a greater risk of false positive detections. In Fig. 4, potential hotspots are highlighted on the PV panel. These may be true PV hotspots or perhaps thermal or infrared radiation or reflection phenomena. This adds complexity and ambiguity to evaluation results. Typically, known faulty PV panels are challenging to photograph in the wild. This scarcity of thermal image data makes supervised ML methods more challenging to develop. As a potential solution, deep learning methods can be combined with traditional computer vision methods to create an improved hybrid method.

We intend to grow this data collection effort to include multi-modal sensor measurements of PV panels. Thermal imagery can be fused with high-resolution standard photographs to maximize pixel resolution, as in Fig. 3. This image data could also be combined with electrical sensor measurements, power generation performance, and information on panel age and wear. We will associate the collected visual data with measurements to create a labeled dataset for supervised ML. We plan to release such a dataset as an ML benchmark for PV panel segmentation, hotspot and defect detection, and electrical performance grading estimation. Commonly, researchers approach PV fault detection as a classification problem: a faulty condition is either present or not. Using computer vision, solar panels could instead be numerically graded for performance, wear, and safety.

Accelerated computing, such as quantum computing, can benefit ML and deep learning methods. We are exploring quantum ML techniques for PV array optimization [34-35].

ACKNOWLEDGMENT

This research project received partial funding from the MRI Award 2019068 and NSF CPS Award 1646542. The ASU SenSIP Center provides the necessary infrastructure for the experiments. Poundra LLC assists in solar testbed research.